\pdfoutput=1

\documentclass[11pt]{article}

\usepackage{acl}
\usepackage{graphicx}
\usepackage{times}
\usepackage{latexsym}

\usepackage[T1]{fontenc}
\usepackage[utf8]{inputenc}

\usepackage{microtype}

\title{CSRCZ: A Dataset About Corporate Social Responsibility in Czech Republic}
\author{Xhesilda Vogli \\ Department of Management \\ Faculty of Economics and Management \\ Czech University of Life Sciences \\ vogli@pef.czu.cz
        \And
        Erion Çano \\ Digital Philology \\ Data Mining and Machine Learning \\ University of Vienna, Austria \\ erion.cano@univie.ac.at}

\begin{document}
\maketitle
\begin{abstract}
As stakeholders’ pressure on corporates for disclosing their corporate social responsibility operations grows, it is crucial to understand how efficient corporate disclosure systems are in bridging the gap between corporate social responsibility reports and their actual practice. Meanwhile, research on corporate social responsibility is still not aligned with the recent data-driven strategies, and little public data are available. This paper aims to describe CSRCZ, a newly created dataset based on disclosure reports from the websites of 1\,000 companies that operate in Czech Republic. Each company was analyzed based on three main parameters: company size, company industry, and company initiatives. We describe the content of the dataset as well as its potential use for future research. We believe that CSRCZ has implications for further research, since it is the first publicly available dataset of its kind.
\end{abstract}

\section{Introduction} \label{sec:intro}

Corporate Social Responsibility (CSR) has evolved from a ``why'' in the early 1950s \cite{Caroll2018} to a ``must'' in recent years. 
Generally, CSR is considered a self-regulating business model which helps companies to contribute to societal goals and be socially accountable to themselves and the public. It is highly influenced by the legal context \cite{https://doi.org/10.1111/jofi.12487} and the socio-political context \cite{Tilt2016Corporate} of the countries where the companies operate. Globally, more and more companies are engaging in CSR initiatives. They are therefore providing more social information to the public. As a result, CSR disclosure has grown to be one of the main study directions for researchers of this field \cite{Goyal2015, Halkosa2016}. 

While reaching adequate standards of sustainability disclosure or reporting is desirable, there are several obstacles to overcome. Sustainability reporting is optional, in contrast, to strictly regulated financial reporting, and it is consequently characterized by a lack of uniformity \cite{Braam2018,Bhattacharyya2015}. Prior studies have been generally focused on the factors that drive the disclosure of these initiatives, the given information, the mode of communication and their impact on the company's performance and image \cite{GONCALVES2023113447,BenoitMoreau2011BuildingBE}. These factors that may influence CSR disclosure reports of a company are usually classified as: (i) \emph{internal}, such as company size, industry sector, financial performance, and corporate governance; (ii) \emph{external}, such as country of origin, stakeholders, media, or social and political environment \cite{Fifka2013CorporateRR,Morhardt2009CorporateSR}. 

Considering the limited research that is available, a few studies also try to investigate the possibility that ``country'' can influence CSR initiatives and disclosure levels \cite{KANSAL2014217,Fufa,KHAN2021100752}. 
On one hand, deeper correlations between other factors and the CSR initiatives of companies are mostly missing. On the other hand, most of the studies (e.g., those cited above) are methodologically ``conservative'' and do not exploit data-driven approaches that have surged in the last decade \cite{en15165775,doi:10.1177/01436244211069655,ccano2018phd}. This trend towards data-driven research is mostly conducted using English language resources (e.g., datasets) which are the most numerous on the internet. There are still several studies and resources in Czech or other languages becoming common and available \cite{nlpinai19,doi:10.1080/09537325.2021.1883583}.   

In this paper, we try to foster data-driven research about CSR by creating and describing CSRCZ, a freely available dataset containing public information of 1\,000 companies operating in the Czech Republic.\footnote{\url{https://zenodo.org/record/7495802}} In the following sections, we present the information retrieval process steps that were followed. We also describe the available data fields (especially those related to CSR), their characteristic values, and some relevant statistics. Finally, we discuss potential utilization of CSRCZ content in the context of future CSR research.   

\section{Dataset Content} \label{sec:datacollection}
The sources for constructing the CSRCZ dataset were collected from the public websites of 1\,000 companies currently operating in the Czech Republic. Initially, the websites of those companies were retrieved by \href{https://www.jobs.cz}{jobs.cz}. Each website was analyzed and only the information relating to CSR was collected. The relevant attributes that were considered are presented in Table~\ref{tab:datasetattributes}.

\emph{Company Name} represents the official name of the company as it is registered in the Czech Republic. It is saved as a text string. \emph{Number of employees} is an integer that includes the total number of full-time employees, part-time employees, seasonal workers, and partners. \emph{Has a CSR page} is a binary value with `1' indicating that this company includes in its website some page with information regarding CSR policies or practices, and `0' indicating that it does not. \emph{Industry Sector} is a string describing the market segment of the company or the type of activity it mostly performs. 

\emph{Size of company} is a categorical variable that describes the size of the company. Any company with fewer than 10 employees is considered as `Micro'. Those with up to 50 employees are `Small' companies. The companies are considered `Medium' if they have 51 up to 250 employees. Any company with 251 or more employees is `Large'. \emph{Initiatives} is probably the most important attribute with respect to the CSR analysis. It is a long string describing any CSR-related policies, practices or initiatives that the company outlines. Finally, \emph{Website} is the URL from which the information was retrieved.       
\begin{table}
\centering
\begingroup
\setlength{\tabcolsep}{15pt}
\begin{tabular}{l c}
\hline
\textbf{Attribute} & \textbf{Content Type}\\
\hline
Company Name & String \\
Number of employees & Integer \\
Has a CSR page & Binary \\
Industry Sector & String \\
Size of company & Categorical \\ 
Initiatives & String \\
Website & URL  \\ 
\hline
\end{tabular}
\endgroup
\caption{\label{tab:datasetattributes}Data attributes and their respective types.}
\label{tab:accents}
\end{table}

\section{Dataset Statistics} \label{sec:datastats}

In the following sections, CSRCZ content is discussed in detail. The characteristics values of the respective fields are analyzed and presented in a tabular format. The codes for deriving the statistics are available online.\footnote{\url{https://github.com/erionc/csrcz-stats}} 

\subsection{Size and Employees} \label{ssec:companysizeemploye}

The size of a company is an important factor that is usually related to the capacities that a company has to implement goals and practices in fulfilment of its CSR strategy. One way to determine the size of a company is by using the number of its employees, same as we described in Section~\ref{sec:datacollection}. This is obviously a simplistic approach, since other factors like different types of assets the company owns (unfortunately, this type of information is not always public) do also indicate how big it is. 

\begin{figure}[ht]
\centering
\includegraphics[width=0.46\textwidth]{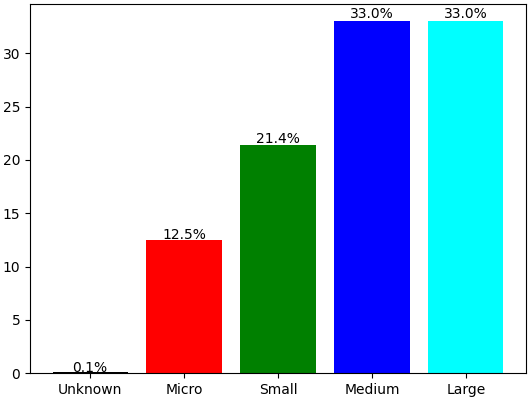}
\caption{\label{fig:companystats}Size distribution of the selected companies.}
\end{figure}

We inspected the collected data and found that most of the companies are large or medium, with each category representing 33\,\% of the instances. There are also 214 small companies which make up 21.4\,\% of the total. There are also 125 companies (representing 12.5\,\% of the total) which are considered to be very small or ``Micro''. For one of the sampled companies, it was not possible to determine its size. The full statistics are presented in Table~\ref{tab:compsizestats} and depicted in Figure~\ref{fig:companystats}. 

\begin{table}
\centering
\begin{tabular}{l c c}
\hline
\textbf{Size} & \textbf{Number} & \textbf{Percent}\\
\hline
Unknown & 1 & 0.1 \\
Micro & 125 & 12.5 \\
Small & 214 & 21.4 \\
Medium & 330 & 33 \\
Large & 330 & 33 \\
\hline
\end{tabular}
\caption{\label{tab:compsizestats}Size statistics of the selected companies}
\end{table}

We also checked the number of employees for each size category. Specifically, we found the minimum, maximum and average number of employees in the `Micro', `Small', `Medium', and the `Large' companies in CSRCZ data. In the case of `Micro' companies, there are at least 5 and at most 9 employees, with an average of 6.54. The same statistics for the case of `Small' companies are 10, 49 and 31.97 respectively. Companies of a `Medium' size have an average of 169.62 employees. Finally, the `Large' companies do have a maximum of 10000 employees (the biggest in CSRCZ) with an average of 1635.58. The statistics are summarized in Table~\ref{tab:employeestats} and depicted in Figure~\ref{tab:employeestats}.      

\begin{table}
\centering
\begingroup
\setlength{\tabcolsep}{10pt}
\begin{tabular}{l c c c}
\hline
\textbf{Company} & \textbf{Min} & \textbf{Max} & \textbf{Avg} \\
\hline
Micro & 5 & 9 & 6.54 \\ 
Small & 10 & 49 & 31.97 \\ 
Medium & 50 & 249 & 169.62 \\ 
Large & 299 & 10000 & 1635.58 \\ 
\hline
\end{tabular}
\endgroup
\caption{\label{tab:employeestats}Minimum, maximum and average number of employees for each company category.}
\end{table}

\begin{figure}[ht]
\centering
\includegraphics[width=0.46\textwidth]{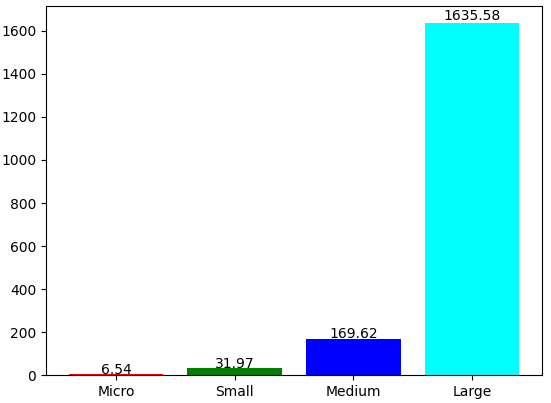}
\caption{\label{fig:employeestats}Average number of employees in each company size category.}
\end{figure}

\subsection{Industry Sector} \label{ssec:companysector}
The industry sector is an interesting attribute since it could shed light on important trends that relate to the CSR initiatives and the different sectors the companies operate. According to GICS (Global Industry Classification Standard), eleven industry sectors represent the majority of industry types nowadays.\footnote{\url{https://www.msci.com/our-solutions/indexes/gics}}  
\begin{description}
    \item[Communication Services] is an industry that includes media and entertainment or any of the telecommunication services.
    \item[Consumer Discretionary] involves the retail industry, hotels, restaurants, leisure, and household durables.
    \item[Consumer Staples] is an industry category that groups all food products, beverages, and tobacco.
    \item[Energy] includes oil, gas, consumable fuels, and energy services.
    \item[Financials] is a category grouping all banking services, capital markets, and insurance services.
    \item[Health Care] involves health care providers and pharmaceuticals.
    \item[Industrials] includes transportation services such as airlines, marine, road \& rail and all services related to it. 
    \item[Information Technology] involves IT services, software, technology hardware, storage, and peripherals.
    \item[Materials] includes all industry sectors that produce chemicals, construction materials, packaging, metals, and mining.
    \item[Real estate] includes real estate investment trusts and real estate services.
    \item[Utilities] includes electric, gas, and water utilities services.
\end{description}

\begin{table}
\centering
\begin{tabular}{l c c}
\hline
\textbf{Sector} & \textbf{Number} & \textbf{Percent}\\
\hline
Unknown & 607 & 60.7 \\
Communication Services & 17 & 1.7 \\
Consumer Discretionary & 91 & 9.1 \\
Consumer Staples & 31 & 3.1 \\
Energy & 15 & 1.5 \\
Financials & 28 & 2.8 \\
Health Care & 16 & 1.6 \\
Industrials & 111 & 11.1 \\
Information Technology & 56 & 5.6 \\
Materials & 25 & 2.5 \\
Real estate & 3 & 0.3 \\
Utilities & 0 & 0 \\
\hline
\end{tabular}
\caption{\label{tab:compsector}Sector statistics of the selected companies}
\end{table}

We explored the data and identified the number and percentage of the companies belonging to each of the above listed industry sectors. The gathered statistics are summarized in Table~\ref{tab:compsector}. Unfortunately, this indicator is not available for many of the data records. Among the available sectors we found, `Industrials' is the most popular, with 111 companies or 11.1\,\% of the total. The sector `Consumer Dicretionary' comes next with 91 companies. `Information Technology', `Cosumer Staples' and `Financials' are also common, with 56, 31 and 28 records each. The most unpopular sectors are `Real estate' and `Utilities', with 3 and 0 companies.   

\subsection{CSR Initiatives} \label{ssec:csrinitiatives}

The most important record attribute of the CSRCZ dataset is probably `Initiatives', where the CSR mission, goals and practices of the companies are summarized. This information usually comes as a sequence of sentences, or sometimes as a few paragraphs. A trivial statistical evaluation here is to check its length in characters or tokens, despite the fact that a short or long `Initiatives' text in the website does not necessarily mean that the CSR commitment of a company is low or high. 

We used NLTK word tokenizer to tokenize the texts.\footnote{\url{https://www.nltk.org/}} Unfortunately, a high number of the sampled companies (more specifically 610 which have 0 length of characters and tokens) have not provided such a description in their websites. The longest CSR initiatives texts have 32023 characters and 4870 tokens. The average length of this attribute is about 1218 characters and 191 tokens. These statistics are summarized in Table~\ref{tab:initiativesstats}. 

\begin{table}
\centering
\begingroup
\setlength{\tabcolsep}{7pt}
\begin{tabular}{l c c c}
\hline
\textbf{CSR Initiatives} & \textbf{Min} & \textbf{Max} & \textbf{Avg} \\
\hline
Characters & 0 & 32023 & 1218.01 \\ 
Tokens & 0 & 4870 & 191.73 \\ 
\hline
\end{tabular}
\endgroup
\caption{\label{tab:initiativesstats}Minimum, maximum and average number of characters and tokens for each CSR initiative.}
\end{table}

\section{Discussion} \label{sec:discussion}

Despite the fact that information is broadly available for a lot of organizations, many companies regularly fail to present the CSR data in a consistent way and assorted according to a framework. As the attention towards CSR is raising and the community becoming more watchful, the need for a standardized definition and CSR framework has been rising. The need for applying data-driven methodologies and providing structured datasets is also in rise. 

The purpose of this work is to foster data-driven CSR research by providing and describing CSRCZ, a recently created dataset. We believe that using CSRCZ can provide a better view of the current understanding of CSR in companies that operate in the Czech Republic and in a global context as well. Various correlations between internal and external company factors and its CSR initiatives can be found. Those findings could be used to develop further frameworks and management strategies in order to better communicate CSR initiatives to stakeholders being those external or internal.
\bibliography{anthology,custom}
\bibliographystyle{acl_natbib}

\end{document}